\documentclass[preprint]{vgtc}               % preprint
\ifpdf%                                % if we use pdflatex
  \pdfoutput=1\relax                   % create PDFs from pdfLaTeX
  \usepackage{graphicx}                % allow us to embed graphics files
  \DeclareGraphicsExtensions{.pdf,.png,.jpg,.jpeg} % for pdflatex we expect .pdf, .png, or .jpg files
\else%                                 % else we use pure latex
  \usepackage{graphicx}                % allow us to embed graphics files
  \DeclareGraphicsExtensions{.eps}     % for pure latex we expect eps files
\fi%

\graphicspath{{figures/}{pictures/}{images/}{./}} % where to search for the images

\usepackage{amsfonts}
\usepackage{microtype}                 % use micro-typography (slightly more compact, better to read)
\PassOptionsToPackage{warn}{textcomp}  % to address font issues with \textrightarrow
\usepackage{textcomp}                  % use better special symbols
\usepackage{mathptmx}                  % use matching math font
\usepackage{times}                     % we use Times as the main font
\usepackage{cite}                      % needed to automatically sort the references
\usepackage{tabu}                      % only used for the table example
\usepackage{booktabs}                  % only used for the table example

\usepackage{color}
\definecolor{mattblue}{RGB}{23,55,180}

\onlineid{1205}

\vgtccategory{Research}

\vgtcinsertpkg

\title{Visually Analyzing Contextualized Embeddings}

\author{Matthew Berger\thanks{e-mail: matthew.berger@vanderbilt.edu}}
\affiliation{\scriptsize Vanderbilt University}

\CCScatlist{
  \CCScatTwelve{Machine Learning}{visual an\-a\-lyt\-ics}{Natural Language Processing}{};
}

\teaser{
 \centering
 \includegraphics[width=.92\linewidth]{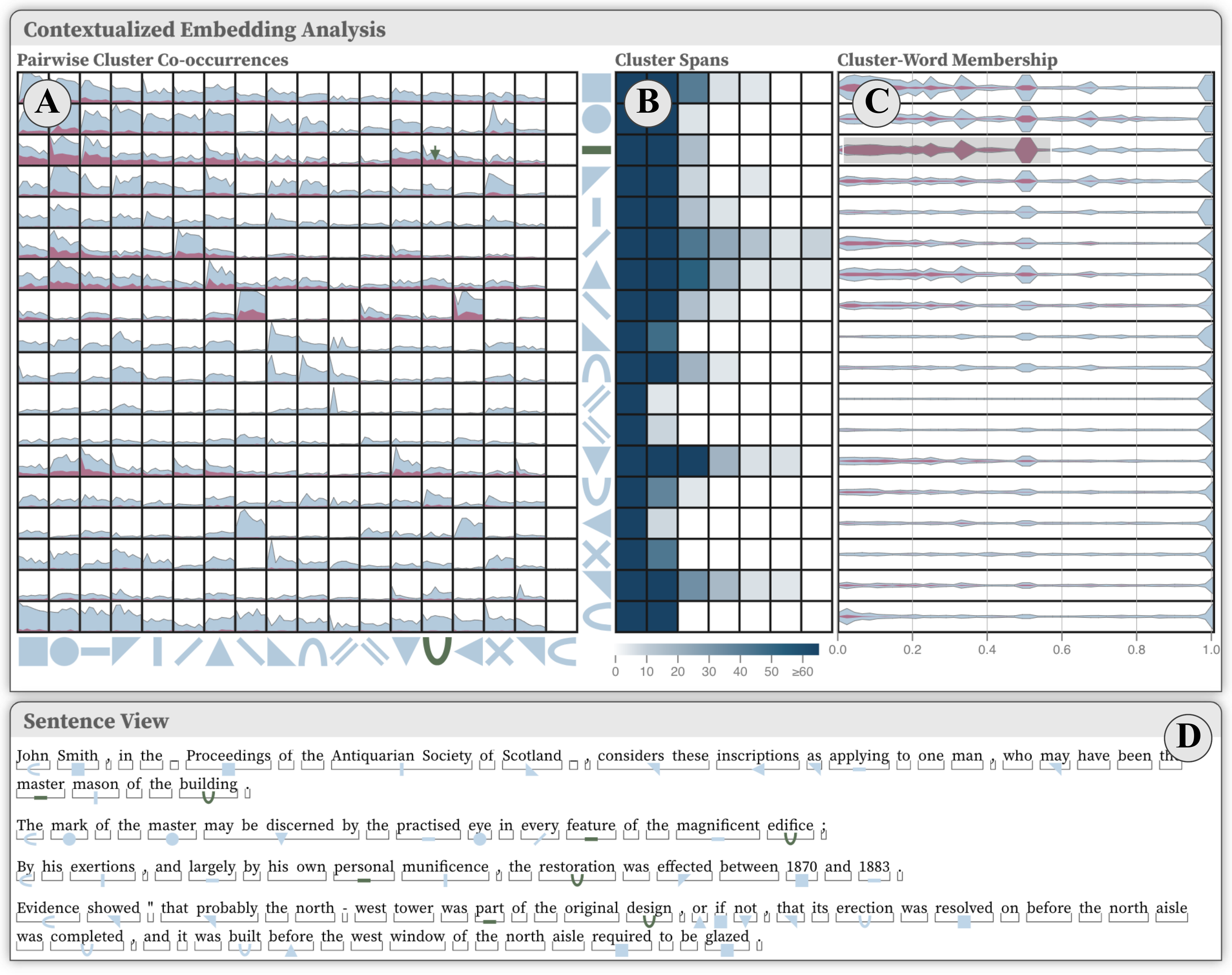}
 \caption{Our method allows the exploration of contextualized embeddings produced by language models. Our design shows (A) co-occurrences of phrases via their assigned clusters, (B) per-cluster span lengths and (C) how much context a given cluster captures. One may also inspect example sentences in detail (D), here highlighting terms that describe building structures.}
 \label{fig:teaser}
}

\abstract{In this paper we introduce a method for visually analyzing contextualized embeddings produced by deep neural network-based language models. Our approach is inspired by linguistic probes for natural language processing, where tasks are designed to probe language models for linguistic structure, such as parts-of-speech and named entities. These approaches are largely confirmatory, however, only enabling a user to test for information known a priori. In this work, we eschew supervised probing tasks, and advocate for unsupervised probes, coupled with visual exploration techniques, to assess what is learned by language models. Specifically, we cluster contextualized embeddings produced from a large text corpus, and introduce a visualization design based on this clustering and textual structure -- cluster co-occurrences, cluster spans, and cluster-word membership -- to help elicit the functionality of, and relationship between, individual clusters. User feedback highlights the benefits of our design in discovering different types of linguistic structures.


} % end of abstract

\begin{document}

\firstsection{Introduction}

\maketitle

% advances in NLP, pre-training, problem: difficult to understand what these models learned
Recent advances in natural language processing (NLP) have led to the development of language models that perform remarkably well across a wide range of language understanding tasks~\cite{peters2018deep,devlin2018bert}, e.g. named entity recognition, entailment, paraphrase verification~\cite{wang2018glue}. These models typically take the form of deep neural networks that are \emph{pre-trained} on a large corpus of unannotated text, and subsequently \emph{fine-tuned} for specific language understanding tasks. An intriguing property of these models is that, due to the combination of the pre-training objective and model capacity, they encode a variety of linguistic structure, despite never being explicitly trained to learn such structure~\cite{clark2019does,liu2019linguistic,reif2019visualizing}. However, comprehending the full space of \emph{what is learned} is elusive, and remains an open problem.

% recent approaches: supervised probes, they suffer from various issues (dataset bias, controlling for model complexity)
Approaches for interpreting pre-trained language models have relied on the design of \emph{supervised probes} -- human-annotated datasets that capture known semantic or syntactic properties, e.g. parts-of-speech, chunking, dependency syntax~\cite{belinkov2019analysis,liu2019linguistic}. Representations extracted from language models are trained to solve problems posed by these probes to assess how well the model captures linguistic structure. Although supervised probes have helped shed light on language models, they inherit several limitations. First, they are confirmatory, only telling us whether or not a language model has learned a known linguistic property. Secondly, models trained to solve probes face issues regarding complexity, e.g.~an overly-complex model that performs well may poorly reflect the probe task~\cite{hewitt2019designing}.

% this work: interactive visualization to understand contextualized embeddings
%   * rather than use supervised probes, instead let the data speak for itself: unsupervised methods, namely clustering
%   * goal of visualization: interpret the clusters
%   * information to extract
%   * visual encodings and interactions to support exploration (reference teaser as appropriate)
In this work we propose an interactive approach to understanding deep, pre-trained, language models. Our work is inspired by existing probing methods, but instead approaches language model interpretability in an unsupervised manner: rather than build probe-specific classifiers, we aim to let the data distribution speak for itself. Specifically, we focus on \emph{contextualized embeddings} of words: vector representations that encode the context of a particular word with respect to its originating sentence. Given a large text corpus, in analogy to supervised probes we \emph{cluster} the embeddings. Given the clustering, the key goal of our visualization is to help a user understand the functionality of clusters, and relationships between clusters. As shown in Fig.~\ref{fig:teaser}, our visualization is designed to highlight patterns of linguistic properties: (A) co-occurrences in clusters, (B) formation of phrases via contiguous cluster spans, (C) just how contextual is a given word, as well as (D) details-on-demand for showing individual sentences and their words' cluster assignments. Combined, these views are designed to help the user identify specific linguistic properties through a set of supported interactions.

% evaluation: hopefully feedback from domain expert!
To evaluate our method we gathered feedback from users to assess what information they could gain by using our system. Through the feedback, we find that different types of linguistic structures, e.g. parts-of-speech, noun phrases, named entities, can be identified through our visualization design.

\section{Related Work}

Our work is most related to interpretability approaches within, both, NLP and visual analytics for understanding language models.

% language models, pre-training (Transformers and BERT)
Neural network-based language models date back to Bengio et al.~\cite{bengio2003neural}, and have gained recent attention with more sophisticated network architectures and language modeling objectives~\cite{peters2018deep,devlin2018bert}. These models have demonstrated significant performance gains in a wide variety of language understanding tasks~\cite{radford2019language,wang2018glue}, despite the seemingly irrelevant tasks used for pre-training, e.g.~masked word prediction and next sentence prediction~\cite{devlin2018bert}. This has motivated the design of supervised probes~\cite{conneau2018you,belinkov2019analysis} as a way to test what linguistic knowledge language models encode in their learned representations~\cite{tenney2018what,liu2019linguistic,jawahar2019does,hewitt2019structural}. Yet these methods face several limitations. As supervised models are usually trained from these representations to assess the accuracy of a probing task, overparameterized models might poorly reflect the linguistic knowledge encoded by the language model~\cite{hewitt2019designing}. Further, it is delicate to design a probing dataset that ensures task relevance in what is learned~\cite{ravichander2020probing,gardner2020evaluating}. Our approach is inspired by probing methods, but is focused on unsupervised methods for interpreting pre-trained language models, complemented by interactive visualization techniques.

% visualization methods for understanding deep learning in NLP
Significant work has been developed within the visual analytics community for interpreting deep NLP models, please see Hohman et al.~\cite{hohman2018visual} for a broader survey on deep learning and visual analytics, and Spinner et al.~\cite{spinner2019explainer} for model interpretability within visualization. Visualization methods have been developed to understand context-independent word embeddings, through assessing analogies~\cite{liu2017visual}, customizing embedding projections~\cite{liu2019latent} and comparing embeddings~\cite{heimerl2018interactive,boggust2019embedding}. Closely related to our method are approaches that visually analyze recurrent neural networks, namely LSTMVis~\cite{strobelt2017lstmvis} and RNNVis~\cite{ming2017understanding}. RNNVis similarly clusters hidden representations of RNNs, but focuses on specific tasks, e.g.~sentiment analysis, whereas we consider task-independent pre-training objectives. Other works have considered the interpretation and editing of sequence-to-sequence models~\cite{strobelt2018s}, models designed for natural language inference~\cite{liu2018nlize}, and interactively performing abstractive summarization~\cite{gehrmann2019visual}. Further methods have visually analyzed self-attention in language models~\cite{vig2019multiscale,park2019sanvis}, whereas we consider contextualized embeddings in Transformer models~\cite{vaswani2017attention}. Recent work such as Checklist~\cite{ribeiro-etal-2020-beyond} and TX-Ray~\cite{rethmeier2019tx} permit the customization of supervised and unsupervised probes, respectively. In contrast to Rethmeier et al., which focuses on interpreting individual neurons, we consider the embedding space as a whole.

\section{Objectives and Tasks}

% approach is inspired by supervised probes for NLP, but seeks a generalization, so that users can identify different types of properties:
%   * linguistic features
%   * semantics (named entities)
%   * relationships (dependencies)
% Tasks:
%   * 

Before discussing the tasks that we aim to support, we first discuss the language model, and extracted representations, used in this work. Our goal is to understand the representations learned by different layers in the Transformer model~\cite{vaswani2017attention}, pre-trained on large amounts of raw textual data using the BERT objectives of masked word, and next sentence, prediction~\cite{devlin2018bert}. Specifically, we use the cased 12-layer BERT model of Devlin et al.~\cite{devlin2018bert}, where for a fixed layer, given a sentence composed of $m$ words $(w_1, w_2, \ldots , w_m)$, passing this sequence through the model provides us with a $d=768$ dimensional vector for each word, denoted $x(w_j) \in \mathbb{R}^d$ for the $j$'th word in the sentence. We denote $x(w_j)$ as the \emph{contextualized embedding} for word $w_j$. Note that the same word's contextualized embeddings from two different sentences will likely be different, due to sentence context, e.g.~``handle'' can be treated as a noun or a verb.

We would like to gain insight on the linguistic properties learned by contextualized embeddings. However, to circumvent the issues inherent in supervised probes, and empower the user in exploration, we approach this in an unsupervised manner. Specifically, given a sentence drawn from a large input corpus, we first obtain the contextualized embedding for each word in the sentence. For a word broken into subwords, the last subword's embedding is taken as the original word's embedding~\cite{liu2019linguistic}. Next, we cluster the contextualized embeddings over all sentences, using k-means. For robustness, we adapt the initialization scheme of Arthur et al.~\cite{plusplus} by limiting seed vectors to unique words, and performing k-means over different initializations, taking the result with lowest sum-of-squared distances to assigned cluster centers. Empirically, we find this scheme produces stable clusters, in part due to the large number of vectors provided by each of our tested corpora~\cite{von2010clustering}, e.g.~ranging from 75K to 250K vectors. We set the number of clusters, $k$, to 50 in all experiments.

For a given sentence $i$ and a word at position $j$ in the sentence, we obtain a cluster label $l(w^i_j) \in [1,k]$. The resulting clustering can be viewed as a \emph{proxy} for a set of supervised probes, e.g.~one cluster could reflect the verb part-of-speech, while another cluster could represent location-based named entities. However, unlike supervised probes, we do not know, a priori, the meaning of the clusters. Hence, the main purpose of our visualization design is to help the user in understanding (1) \emph{what a cluster represents}, and (2) the \emph{relationships between clusters}. The tasks supported in our design aim to address these objectives, and serve to \emph{abstract} typical approaches taken in supervised probes:

(\textbf{T1}) \textbf{Assess how much context a given cluster contains.} Certain words (e.g.~punctuation) are less reliant on context than other words (e.g.~``place'') that may have multiple senses. This task intends to abstract multiple probes such as parts-of-speech~\cite{liu2019linguistic}, semantic role labeling~\cite{bjerva2016semantic}, and word tense~\cite{conneau2018you}.

(\textbf{T2}) \textbf{Determine a cluster's ability to form meaningful phrases.} This task abstracts segmentation probes such as syntactic chunking and named entity extraction~\cite{liu2019linguistic}, as well as constituency parsing~\cite{kim2019pre}.

(\textbf{T3}) \textbf{Analyze relationships between clusters.} This task abstracts relationships between clusters, e.g.~relation extraction~\cite{liu2019linguistic}, syntactic dependencies~\cite{clark2019does}, and coreference resolution~\cite{tenney2018what}.

\section{Visualization Design}

\begin{figure}[t]
\centering
\includegraphics[width=1\linewidth]{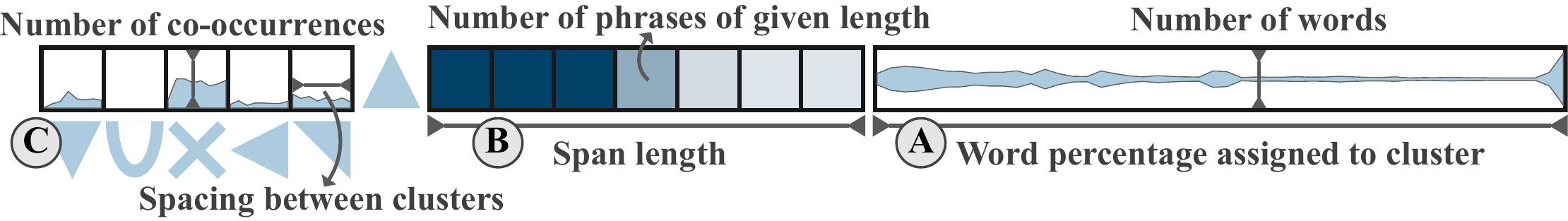}
\caption{Overview of our design showing (A) relative amount of context encoded by a cluster's set of words (B) frequency over different span lengths for cluster-specific words forming contiguous spans, and (C) cluster co-occurrence frequency regarding word/phrase spacing.}
\label{fig:design}
\end{figure}

In this section we discuss our visualization design that addresses our tasks, please see Fig.~\ref{fig:design} for an overview of the encodings employed in our design.

\subsection{Cluster-Word Membership}

This view address (\textbf{T1}) in showing the amount of context reflected in a given cluster. Specifically, for a given word $w$, denote $c(w,l)$ as the number of times this word appears in the corpus with cluster $l \in [1,k]$. Then for such a cluster, we compute the percentage in which that word appears in the cluster:
\begin{equation}
p(w,l) = \frac{c(w,l)}{\sum_{j=1}^k c(w,j)}.
\end{equation}
Thus, for cluster $l$ we have an assigned percentage $p$ for all words $w$ in our corpus. We encode this as a distribution (Fig.~\ref{fig:design}(A)), where the x-axis encodes the percentage, and an area mark's height encodes how many words contain that percentage. We perform kernel density estimation to arrive at a smoothed distribution. Percentages of $p = 0$ are filtered out, as they tend to dominate, and are implicitly encoded via nonzero counts across the rest of the clusters. This view enables us to determine differences in clusters in terms of word senses. For instance, two clusters may both reflect past tense, yet they are distinguished by part-of-speech, where one cluster represents adjectives, and the other represents verbs. Our design would consequently depict overlap between these clusters (\textbf{T1}). In general, distributions that are concentrated at a value of 1 indicate only one meaning, independent of context, whereas a more even distribution across percentages indicates the dependence on context for the meaning of individual words.

\subsection{Cluster Spans}

This view addresses (\textbf{T2}) in showing the ability of a cluster to represent contiguous text spans. Specifically, for a given cluster, for each sentence in our corpus we group words that (a) form a contiguous span and (b) all belong to this particular cluster. We then count how many times, for a given span length, these cluster-specific spans occur over the entire corpus. We visually encode this as a heatmap (Fig.~\ref{fig:design}(B)) where each square represents a particular span length, beginning at a span of 1 (individual word), and increasing from left-to-right. We use a sequential, luminance-decreasing color map to encode count, e.g.~how many times a cluster-specific span occurs in the corpus. Aligned columns of the heatmap permit a rapid comparison of span length frequencies between clusters, while a given row depicts a cluster's distribution of span frequencies. As shown in Fig.~\ref{fig:teaser}(B) for the last layer of the Transformer~\cite{vaswani2017attention} model, this design enables the user to quickly assess whether certain clusters result in long spans compared to other clusters, indicative of certain types of linguistic features, e.g.~named entities or a part of a constituency parse tree. This grouping of words into contiguous, cluster-specific spans is carried over to other elements of the design, namely cluster co-occurrences, as well as the detailed sentence view. Herein we refer to these grouped words as \emph{phrases} for full generality.

\subsection{Pairwise Cluster Co-occurrences}

This view addresses (\textbf{T3}) in depicting relationships between clusters. More specifically, for a given phrase corresponding to a cluster, we count how many times it co-occurs with a different cluster's phrase within a given sentence. We measure co-occurrences over different spacings of phrases, e.g.~phrases belonging to two different clusters might be right next to each other, but other times they might be separated by several phrases.  We show these relationships in a small-multiples view of area marks: rows correspond to clusters in the first position (e.g.~the left portion of the co-occurrence), while columns correspond to clusters in the second position (e.g.~the right portion). The height of the area mark encodes the number of co-occurrences, while the x-axis within each cell encodes the amount of spacing between phrases, increasing from left-to-right (Fig.~\ref{fig:design}(C)). Area marks allow the user to quickly identify patterns with respect to cluster pairings. A large spike within the area mark indicates a frequent co-occurrence between clusters at a given amount of spacing, distinguished from other spacings between these clusters. This potentially indicates a salient relationship between clusters (\textbf{T3}), e.g.~co-reference resolution for diagonal cells (identical clusters) or dependency relations between distinct parts-of-speech spaced a fixed amount apart. Note in Fig.~\ref{fig:teaser}, there are zero counts for co-occurrences that are directly next to each other in cells on the diagonal, due to the grouping of words into phrases.

Further, to visually align the different views, we associate a unique glyph with each cluster, distributed as horizontal and vertical spans within the co-occurrence view. In particular, the vertical strip of glyphs is in alignment with the rows of the span heatmap and cluster-word membership views, for quick identification of clusters amongst all views. This glyph design was chosen to handle a potentially large number of clusters. Other visual channels, e.g.~color, can lead to discriminability issues, particularly for complex spatial arrangements~\cite{haroz2012capacity}. This is characteristic of our sentence view, discussed next.

\begin{figure}[t]
\centering
\includegraphics[width=0.94\linewidth]{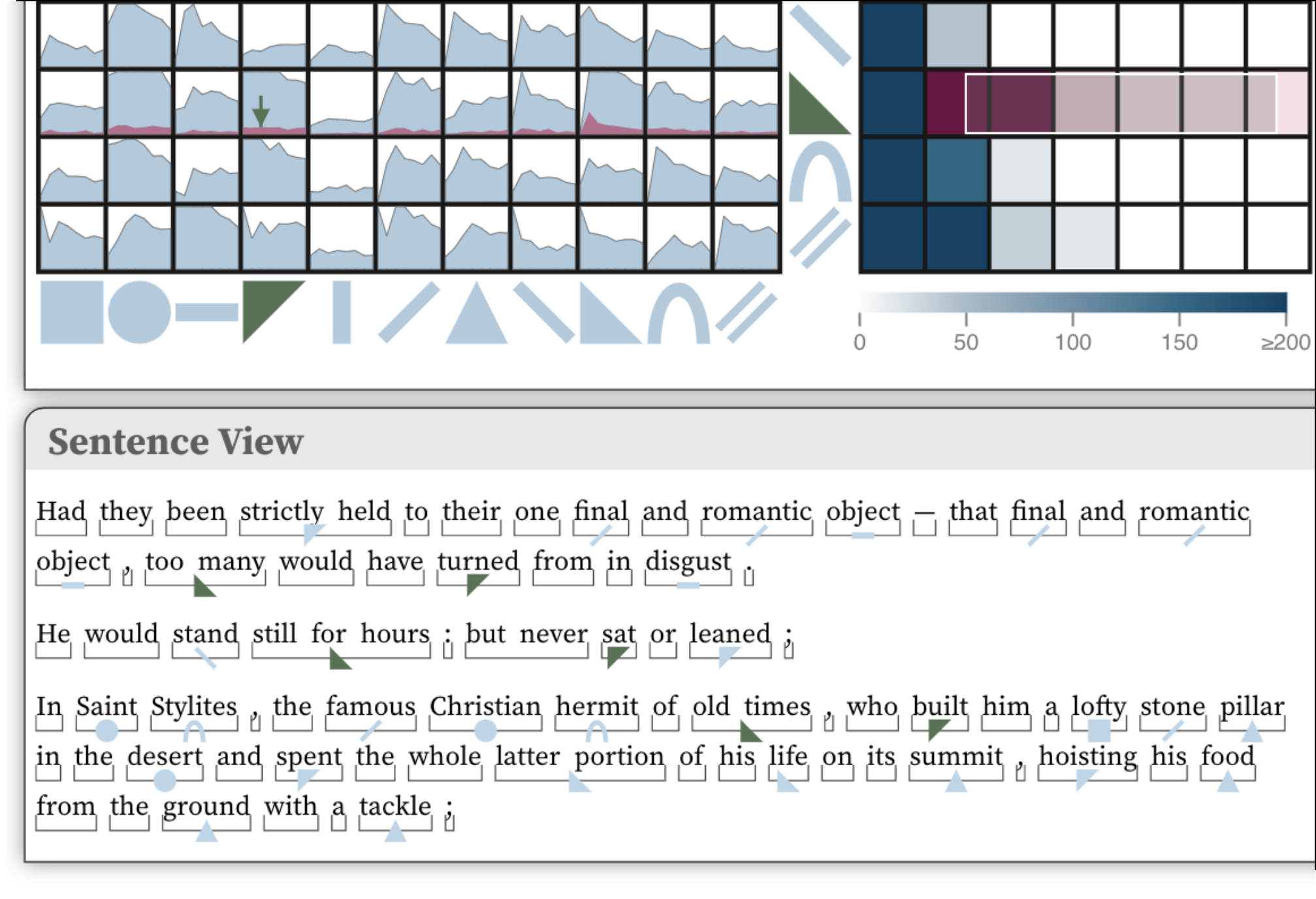}
\caption{Here we show how the user can discover multi-word phrases of the concept of time through our interface. Brushing spans of length greater than 1, and selecting in the co-occurrence view, we obtain detailed inspections in the sentence view that enables this discovery.} 
\label{fig:spans}
\end{figure}

\subsection{Interactions and Detailed Sentence Inspection}

\begin{figure*}[t]
\centering
\includegraphics[width=1\linewidth]{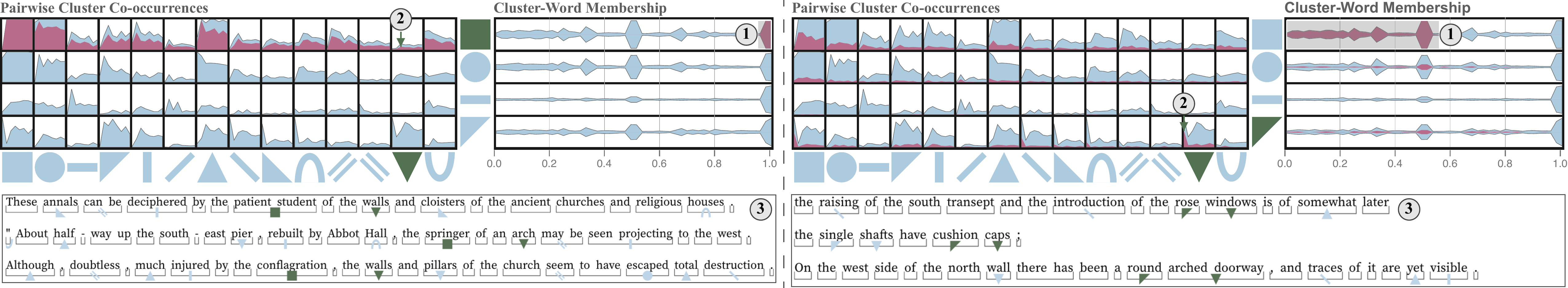}
\caption{We show a use of our system for exploring multiple word senses, allowing the user to discover nouns that are largely context-independent (\textbf{left}), in contrast to context-dependent words shared by a different cluster that capture adjectives (\textbf{right}).} 
\label{fig:usecase}
\end{figure*}

We allow for user interactions to (a) understand relationships between the different views, and (b) provide for detailed inspection of sentences. Specifically, the user can brush the cluster-word membership view to select words within the particular percentage range. The remaining cluster-word distributions are updated for the brushed set of words with a superimposed purple area mark, in order to show more detailed relationships between clusters, shown in Fig.~\ref{fig:teaser}(C). Furthermore, the co-occurrence view is also updated, where we show a purple area mark for co-occurrences that contain the brushed words. We limit this filtering only to the first item (left position) of a co-occurrence. This linked update allows the user to inspect co-occurrences that have varying levels of context, depending on the user's selection. We, similarly, allow the user to brush the cluster span view, limiting phrases to the particular span lengths brushed by the user. We, further, update the co-occurrences view to this filtered set of (left positional) phrases, but limit the selection to \emph{only} the particular cluster, in contrast with the word-cluster membership selection which impacts \emph{all} clusters.

We also allow for the user to select both an \emph{individual cell} and \emph{phrase spacing} within the co-occurrence view, as indicated by the dark green arrow in Fig.~\ref{fig:teaser}(A), and corresponding highlighted cluster glyphs. If a user has previously performed a brushed from the aforementioned interactions, then this selection is limited to the brush query: this is shown in Fig.~\ref{fig:teaser}(A) by the arrow positioned on the purple area mark. For a given selection, we populate a more detailed sentence view in Fig.~\ref{fig:teaser}(D), where we show sentences that contain the particular pair of clusters, and spacing between clusters. The cluster-specific glyphs are carried over to this view, as well as the depiction of phrases via brackets that highlight cluster-specific contiguous spans. In Fig.~\ref{fig:teaser}(D), we see that the user's selection resulted in, predominantly, adjectives in the left cluster, yet these are words that can have different senses (e.g.~``master'' can be an adjective or noun), which arise from the user's brush of words that belong to different clusters, and are thus more context-dependent. Likewise, Fig.~\ref{fig:spans} shows an example of filtering phrases to within a certain length.

% other inputs
We allow the user to control various aspects of the design. They may select any of the layers within the Transformer model to load in the main view, providing a quick comparison of how contextual particular layers are -- including the first layer, which is largely dependent on word embeddings and thus mostly free of context. The user can also control how many clusters to show in the visualization to reduce visual complexity, where we prioritize clusters based on the number of unique words that each cluster contains. Further, for the sentence view the user can opt to exclude glyphs of clusters not selected in the co-occurrence view, freeing clutter.

\section{Results}

To demonstrate our interface, we first show a use case of our system. Our interface supports the loading of an arbitrary set of sentences, but for evaluation purposes, we limit this to sentences from a book, namely ``Scottish Cathedrals and Abbeys.''~\footnote{texts from books are acquired from Project Gutenberg (\url{https://www.gutenberg.org/}). Though the main results shown are based on only one book, Fig.~\ref{fig:spans} shows our interface for ``Moby Dick''.} Our use case is based on this corpus, showing results for the 9'th layer of the Transformer model, please see Fig.~\ref{fig:usecase}. On the left side, the user first selects words that have high membership with the square cluster (1), thus limiting our view to context-independent words. The selection prompts an update to the co-occurrence view via the purple area marks representing those words, where upon clicking a pair of clusters (2) we see that the square cluster for this selection reflects nouns (3). On the right, the user next selects a range of word-cluster memberships from the same (square) cluster (1) prompting linked highlighting across clusters, thus reflective of context-dependent words/phrases. We can observe a spike in co-occurrences for this selection with respect to a pair of clusters (2), indicative of words that belong to different clusters that are right next to one another. Upon closer inspection (3), we find that this represents adjective-noun pairs, where the words classified as adjectives may also be treated as nouns, demonstrating their reliance on context.

% describe data and participants
In addition, we have gathered feedback from users, in order to assess what features participants could find by interacting with the visualization. More specifically, we conducted experiments with three graduate students, all in Computer Science, who all have some amount of experience using visual interfaces. We did not constrain them in their interactions, instead promoting free-form exploration, asking them: (1) What insights did you find by using the interface? (2) Did you find the interface easy to use?

All in all, participants found different aspects of language through the interface: one participant was able to quickly identify parts-of-speech (adjectives, nouns), while another participant found named entities in the form of dates, as well as more semantic groupings, e.g.~different aspects of religion such as church, chapel, etc.. and building structures such as monument, exterior, etc.. Another participant was able to discover patterns with respect to different layers of the Transformer model, namely how the cluster-word membership becomes less unique in later layers, as well as more diverse span lengths.
Participants, however, did find the design to be rather complex. One participant mentioned that it took some time to understand it, but afterwards, they were able to navigate amongst the views. Another participant, however, found the complexity to be too overwhelming at times, which inhibited their discovery.

\section{Conclusion}

We introduced a method for visually analyzing contextualized embeddings produced from deep, pre-trained, language models. Our visualization design takes inspiration from, and abstracts, the class of supervised probes traditionally used to interpret language models, in order to enable a more general analysis of contextualized embeddings. We find preliminary user feedback to be encouraging, however, in the future we plan on obtaining feedback from domain experts within NLP as well as linguistics to assess the design's effectiveness. Furthermore, we plan on extending our work to enable a more comparative analysis of contextualized embeddings, particularly across layers, in order to understand what linguistic properties are learned amongst different representations.

% summarize work; future: other ways to compare embeddings, and more detailed user study from linguistics/NLP experts

%\bibliographystyle{abbrv}
\bibliographystyle{abbrv-doi}

\bibliography{template}
\end{document}